\documentclass[preprint,showpacs,groupedaddress]{revtex4}
\usepackage[greek,english]{babel}
\usepackage{graphicx}
\usepackage{amsmath}
\usepackage{mathrsfs}
\usepackage{amsfonts}
\usepackage{amssymb}
\usepackage[utf8]{inputenc}

\begin{document}

\title{Steering in-plane shear waves with inertial resonators in platonic crystals}

\author{Younes Achaoui}
\affiliation{Aix-Marseille Université, CNRS, Centrale Marseille, Institut Fresnel, UMR 7249, 13013 Marseille, France}
\author{André Diatta}
\affiliation{Aix-Marseille Université, CNRS, Centrale Marseille, Institut Fresnel, UMR 7249, 13013 Marseille, France}
\author{Sébastien Guenneau}
\affiliation{Aix-Marseille Université, CNRS, Centrale Marseille, Institut Fresnel, UMR 7249, 13013 Marseille, France}

\begin{abstract}

Numerical simulations shed light on control of shear elastic wave propagation in plates structured with inertial resonators. The structural element is composed of a heavy core connected to the main freestanding plate through tiny ligaments. It is shown that such a configuration exhibits a complete band gap in the low frequency regime. As a byproduct, we further describe the asymmetric twisting vibration of a single scatterer via modal analysis, dispersion and transmission loss. This might pave the way to new functionalities such as focusing and self-collimation in elastic plates. 

\end{abstract}


\maketitle

Over the late nineties, structured materials have attracted particular attention due to their functionalities regarding waves' control. Regardless of the wave vector, the direction of propagation and polarisation, phononic crystals and locally resonant sonic crystals can prohibit the propagation of elastic waves by means of destructive interferences due to Bragg scattering and local resonances of each individual scatterer, respectively~\cite{kushwahaPRL1993,shengSc2000} . Moreover, waves can easily be channelled through waveguides~\cite{khelifPRB2003}, slowed down or trapped to enhance interaction between photons and phonons for example~\cite{maldovanAPB2006}. Furthermore, locally resonant sonic crystals can be viewed as acoustic metamaterials, counterparts of electromagnetic metamaterials, which were introduced by Pendry and his colleagues to enhance coupling between resonators~\cite{pendryIEEE1999}.

The first studies were dedicated to bulk waves where no supplementary boundary condition were taken into account. The concepts discussed above have been quickly transposed to surface guided waves and Lamb waves~\cite{meseguerPRB1999, khelifPRE2006}. In 1998, Tanaka and Tamura calculated the band structures to investigate the dispersion of surface elastic waves in a two dimensional semi-infinite phononic crystal~\cite{tanakaPRB1999}. The directional and complete surface band gaps in drilled-hole phononic crystals have been presented by Wu et al. and Benchabane et al., respectively~\cite{wuJAP2005, benchabanePRE2006}. The locally resonant mechanism was compared to Bragg scattering for surface elastic waves a few years later~\cite{khelifPRB2010}. 

Unlike for surface elastic waves where very few studies have been conducted, many papers were published on waves propagating in structured plates; especially flexural waves where both analytical approximations (Kirchhoff, Mindelin) and exact numerical solutions were reported. The zero frequency band gaps in arrays of infinite conducting wires~\cite{mcphedran} and in pinned plates~\cite{craster}, lensing in platonic crystals~\cite{duboisAPL2013} or low frequency band gaps in locally resonant sonic crystals with stiff pillars~\cite{wuAPL2008, pennecPRB2008}, heavy pillars combined with soft rubber~\cite{oudichNJP2010} or Helmholtz-like resonators~\cite{hsuJPd2011} are examples of those. 

\begin{figure}[!t]
\centering
\includegraphics[clip,angle=0,width=140mm]{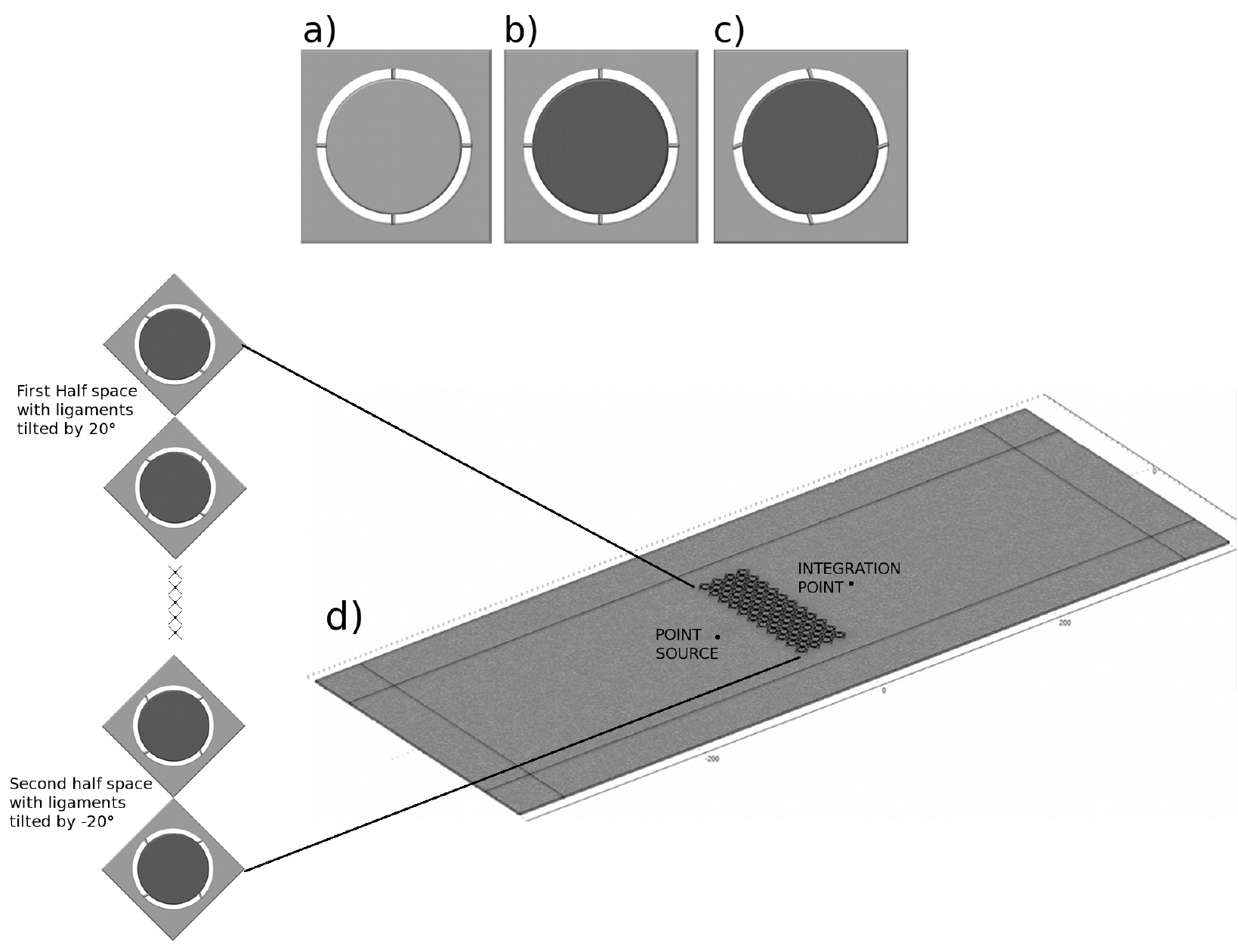}
\caption{
Schematics of three configurations used for investigating the propagation of shear phonons in a periodic array of inertial resonators linked to a $2$~mm-thick plate through short tiny ligaments. a) Both the plate and the resonators are made of duraluminium and the ligaments are aligned. b) The central inclusion is made of tungsten (heavy material) and connected to the host plate through aligned ligaments. c) Same as b) but now with ligaments tilted by 20~degrees. d) Schematics of the periodic structure used to achieve collimation and focussing. The central inclusions are made of tungsten and connected with ligaments tilted by 20~degrees~/~-20~degrees in order to facilitate the change of the direction of the wave propagation at the resonance frequency.  
}
\label{crystal}
\end{figure}

\begin{figure}[!t]
\centering
\includegraphics[clip,angle=0,width=87mm]{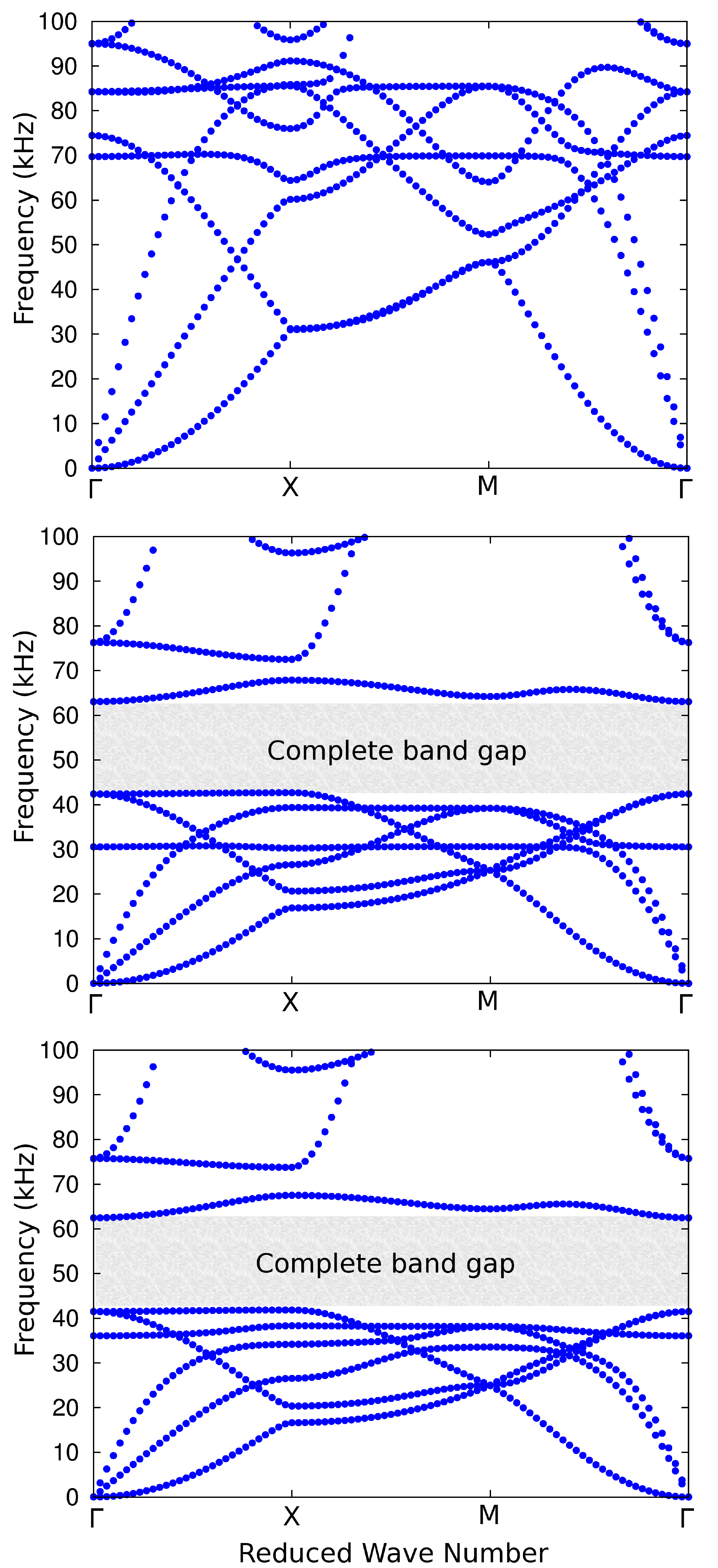}
\caption{
Band diagram of elastic wave propagation within platonic crystals with inertial resonators corresponding to the three configurations in Fig. 1(a) (upper panel), Fig. 1(b) (middle panel) and Fig. 1(c) (lower panel). $\Gamma$, X and M denote the vertices $(0,0)$, $(\pi/d,0)$ and $(\pi/d,\pi/d)$ of the first Brillouin zone, where $d$ is the pitch of the array.
}
\label{dispersion}
\end{figure}

In the case of plates with resonators, many researchers explored different configurations to engineer band structures by tailoring the overall shapes of the device. For instance, Assouar et al. ~\cite{assouar} and Wang et al. ~\cite{wang} studied numerically the enlargement of locally resonant acoustic band gaps and the propagation of Lamb waves in two-dimensional sonic crystals based on a double-sided stubbed plate. Wang et al. implemented the optimized design of alternate-hole-defect on a silicon PC slab in a square lattice. Recently, Zhang et al. introduced a spiral-like resonator as a unit cell to obtain large band gaps in the low frequency range~\cite{zhang}.  

In this letter, we describe a structure composed of inclusions, heavier than the host plate, sustained by very tiny ligaments. We first demonstrate that such a configuration enables a strong attenuation of every elastic wave polarisation in a wide frequency range compared to its equivalent monolithic counterpart. We present afterwards the asymmetric twisting vibration mechanism induced by tilted ligaments and its aptitude to focus elastic shear waves. Such structures have been reported previously in the case of bulk waves but were limited to the demonstration of band gaps ~\cite{bigoniPRB2013, bertoldiPRL2014}.

The unit cell of the model considered here is a $3.5$~mm radius cylindrical inclusion embedded in a perforated plate, sustained through tiny ligaments. Inclusions, ligaments and the host plate are all $2$~mm thick. We compare three configurations, namely, the monolithic duraluminium plate with aligned ligaments, tungsten inclusions in a duraluminium plate with aligned ligaments and finally, tungsten inclusions in a duraluminium plate with ligaments tilted by 20~degrees (from top to bottom panels in figure \ref{dispersion}). Since the whole structure is of constant thickness, the plate could be easily incorporated in plate devices and is a good candidate to replace freestanding membranes. The physical parameters used in these simulations are $\rho_{tun}~=~17800$~kg/m$^3$, $\lambda_{tun}~=~166$~GPa and $\mu_{tun}~=~129$~GPa for tungsten and $\rho_{al}~=~2790$~kg/m$^3$, $\lambda_{al}~=~56.2$~GPa and $\mu_{al}~=~28.7$~GPa for aluminium. We first perform the band structure calculation in the first Brillouin zone by imposing Floquet-Bloch conditions on opposite sides of the unit cell, see Fig. \ref{dispersion}. In figure \ref{dispersion}-a), it is shown that the presence of the resonators does not disturb the wave propagation up to $60$~kHz. The frequency range [$60$~kHz, $100$~kHz] is characterized by the presence of coupling between the inclusions' resonance and in-plane coupled pressure and shear waves. The resonance frequency observed at $70$~kHz of the twisting mode can be estimated by the formula~\cite{guenneau2007} :

\begin{equation}
f = \frac{1}{2\pi}\sqrt[] {\frac{N(\lambda+\mu)\mu}{3(\lambda+2\mu)M} \sum_{i=1}^N {\frac{\epsilon ^3 h^3}{(b_i -a_i)^3}}}
\end{equation}

\begin{figure}[!t]
\begin{center}
\includegraphics[clip,angle=0,width=85 mm]{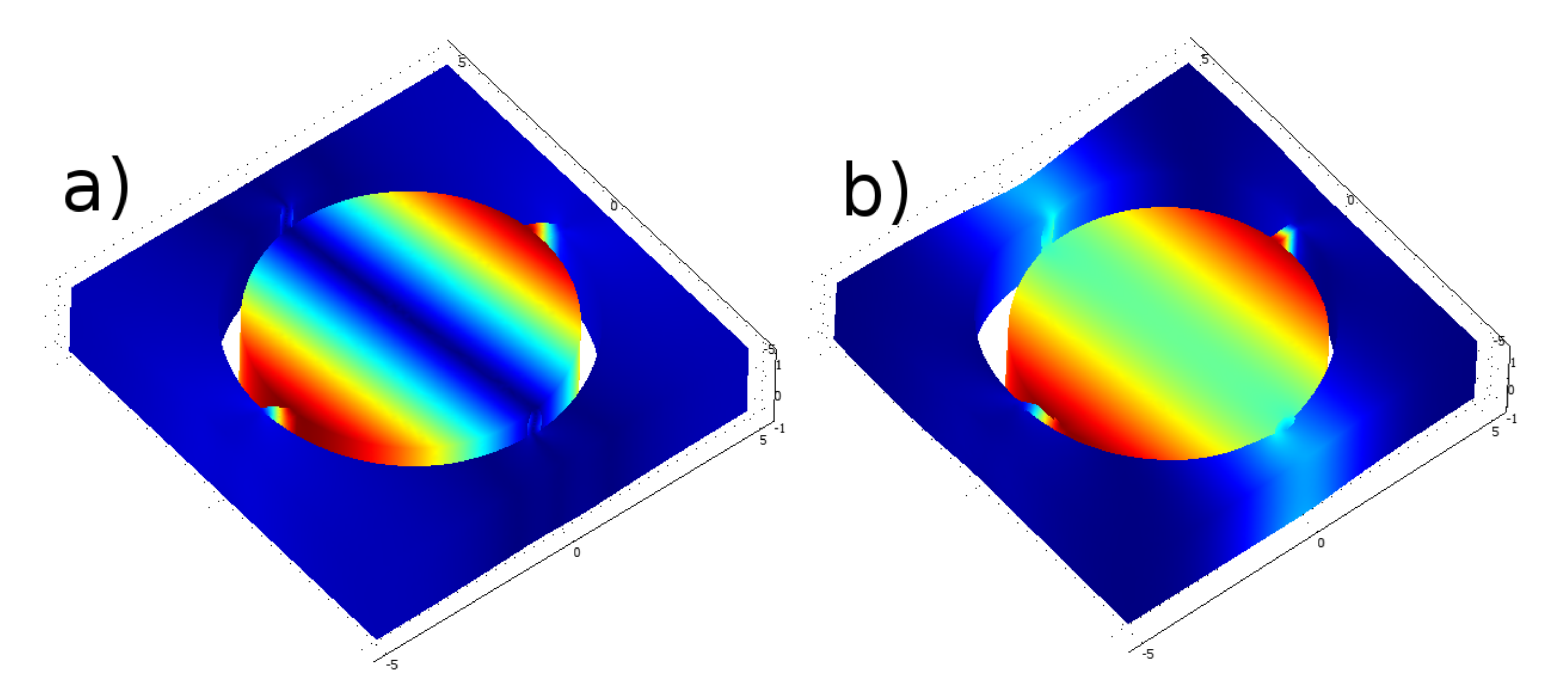}
\end{center}
\caption{Illustration of the twisting modes (a- at $30$~kHz, b- at $32$~kHz ) in the case of aligned ligaments and ligaments titled by $20$~degrees of tungsten-inclusion-based platonic crystal. It highlights the energy transfer from the twisting mode to the plate. The reader may refer to supplementary materials for better viewing. }
\label{eigenmode}
\end{figure}

Here $M~=~R^2\pi~\rho$, denotes the mass of the inner resonator and R being the inner radius, $\epsilon~h~=~0.2$~mm the ligaments width, $b_i-a_i~=~0.5$~mm the ligaments length and $N~=~4$ the number of ligaments involved. When the mass is connected to the plate through one ligament, both the calculated and the computed resonance frequencies are equal to $37$~Hz. It should be noted that in the above formula $\rho$ has the dimension of kg/m$^2$ since in-plane waves do not sense the third direction. Unlike the monolithic plate where both the ligaments and the inner mass are made of the same material as the host plate, this formula fails to estimate the resonance frequency when different materials are considered. For the rest of the document, only results based on finite element numerical simulations are presented. The inner duraluminium resonator in the unit cell has been replaced afterwards by a tungsten one, the same geometrical properties hold (figure \ref{crystal}-b). In figure \ref{dispersion}-b, we depict the dispersion curves of this structure. It is shown that a complete band gap occurs between $40$~kHz and $60$~kHz. Moreover, one can notice the presence of a flat band at $30$~kHz. This corresponds to a rotational mode, difficult to couple with the continuum because of the heavy core resonator. In order to facilitate this inertial resonance, we tilt each ligament by twenty degrees. In figure \ref{dispersion}-c) we draw the dispersion curves for such a structure. It is worth noting that these bands match nearly perfectly the dispersion curves obtained in the case of aligned ligaments (figure \ref{dispersion}-b)) except near the resonance between $30$~kHz and $40$~kHz. Indeed, a strong coupling between the rotational mode on the one hand, the shear wave and the compressional wave on the other hand, is suggested by numerous level repulsions between bands. We further illustrate this hybridization between the continuum and the local resonance of the inclusion in figure \ref{eigenmode}. It can be easily seen that the resonator in the basic cell twists independently without inducing any collateral vibration of the host plate in the case of aligned ligaments. By contrast, the resonator communicates more energy to the host plate when the ligaments are tilted. The phenomenon is best seen with movies of the eigenmodes' vibration (see supplementary materials)~\cite{SM}. This strong coupling between the host plate and the resonators is accompanied by a redirection of shear waves in a privileged direction caused by the intrinsic asymmetric shape of the resonators. This observation can be made regardless of the direction of propagation. This makes these kinds of resonators good candidates for focusing, collimation and prohibition of waves, as we shall now explain.

\begin{figure*}[!t]
\centering
\includegraphics[clip,angle=0,width=170mm]{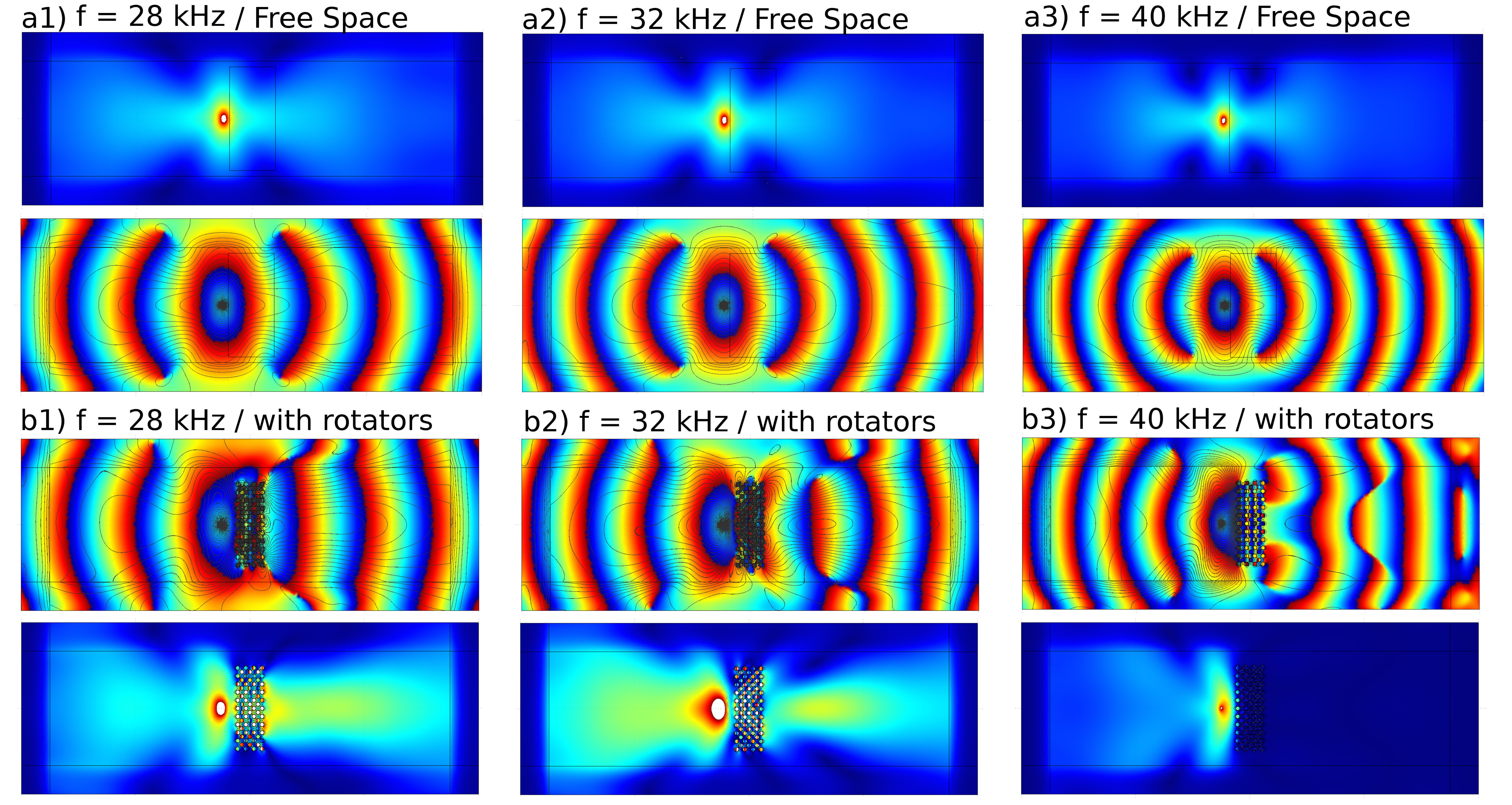}
\caption{Energy cartography, induced phase shift and iso-surfaces of shear in-plane wave in the presence or absence of a seven-column platonic crystal made of rotators with tilted ligaments. It underlines different functionalities that may be achieved at different frequencies. a) Endoscope effect, b) focusing and c) total mirror reflection.
}
\label{lensing}
\end{figure*}

\begin{figure}[!t]
\begin{center}
\includegraphics[clip,angle=0,width=150 mm]{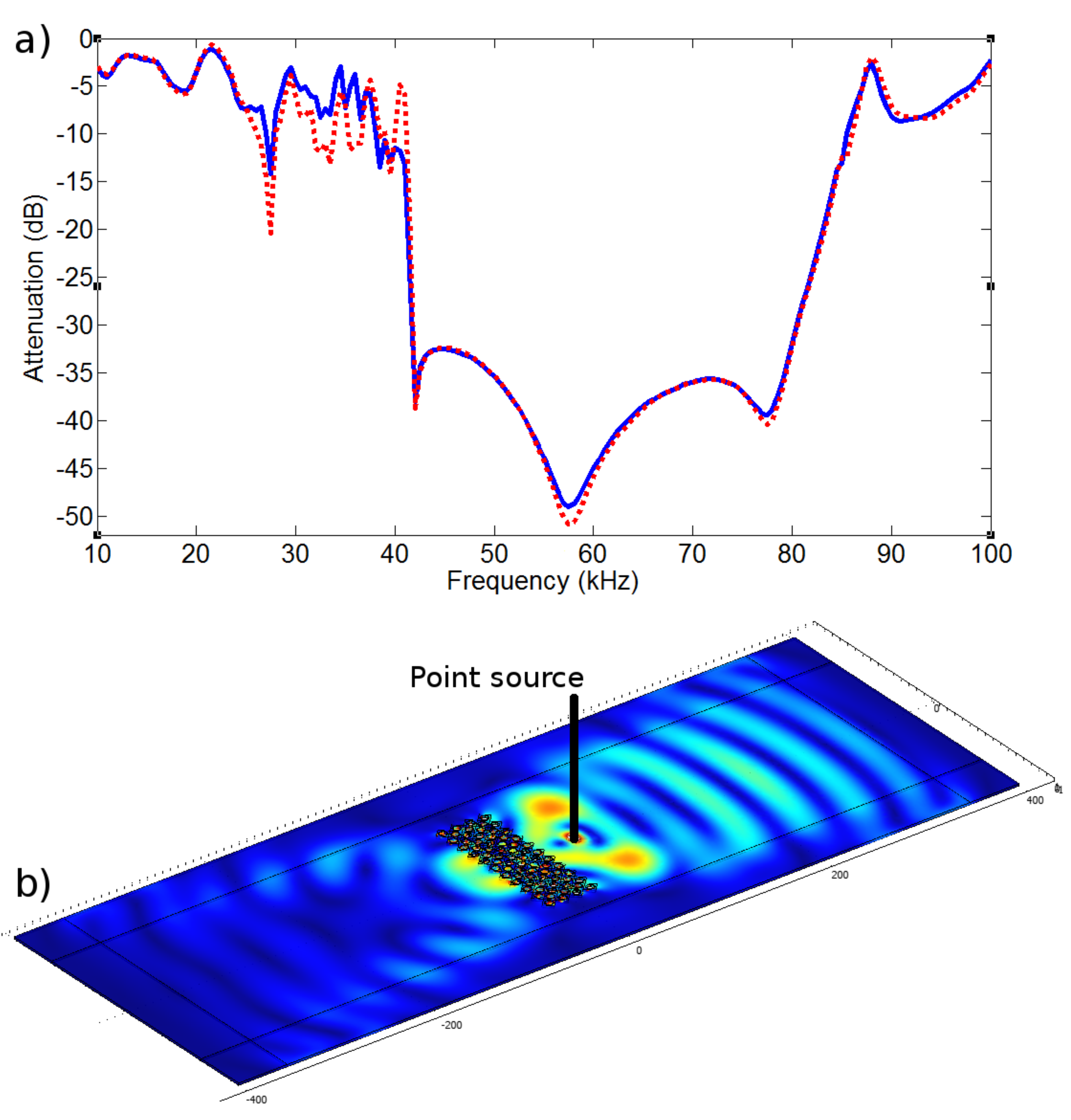}
\end{center}
\caption{a) Transmission loss through a seven-column platonic crystal made of rotators with tilted ligaments. The solid blue curve describes the case when the rotators are placed in the right direction to convey the elastic waves and the dotted red when the rotators counteract its energy flow. The two curves fit perfectly except for frequencies close to resonance. b) Example of crystal response with counteraction effect at $32$~kHz to be compared with Figure 4-b2.}
\label{transmission}
\end{figure}

Figure \ref{lensing} summarizes the aforementioned functionalities. In each sub-figure, we plot the magnitude of the field, its phase shift and its iso-surfaces in order to highlight physical phenomena at different frequencies. In figures \ref{lensing}-a and \ref{lensing}-b we depict the propagation of the shear elastic wave in the homogeneous plate and in the presence of the periodic structure, respectively. The spatial energy repartition and the cylindrical phase shift in the homogeneous plate express the consequence of a point source generating shear vibrations. The non-dispersive feature of shear waves makes them travel with the same group and phase velocities and hence maintains the same behaviour regardless of frequency. The energy isosurfaces allow us to follow the energy decay in the homogeneous plate. In figure \ref{lensing}-b1, in the presence of the periodic structure, the $28$~kHz-frequency corresponds to the directional band gap for shear waves (second mode in figure \ref{dispersion}-c). This leads to a self collimation (or endoscope effect). Through the phase shift analysis, we notice that the curvature of wavefronts is reduced thanks to the periodic structure and the energy isosurfaces are parallel over a longer distance. The energy repartition corroborates this observation. The most interesting effect can be seen at frequency $32$~kHz. Indeed, this frequency represents the aforementioned inertial twisting mode (figure \ref{eigenmode}-b). The sub-wavelength resonators convey collectively the energy towards the point source image. At the crystal exit, we can clearly notice the change in wavefront curvature (from convex to concave) and the concentric energy isosurfaces corroborated by the magnitude cartography. Finally, we evaluated the well-known perfect mirror effect caused by the complete band gap for shear elastic waves. 

In figure \ref{transmission}-a, we quantify the energy loss through the periodic structure collected from a sensor placed on the opposite side to the point source. The numerical simulations have been performed in both situations, when the point source is placed on the left side (red curve) and the right side (blue curve) of the periodic structure. One can first notice that the two curves retrace perfectly the position of the dispersion-based complete band gap. Furthermore, the red curve fits entirely the blue one except in the resonance region between $30$ and $40$~kHz. We were particularly interested in the energy steering at the resonance frequency. In contrast to figure \ref{lensing}-b2, the rotators' movements counteract the energy flow and destroy completely the wave's fringes (figure 5-b).

In conclusion, we analysed the propagation of elastic waves in a platonic crystal with inertial resonators. It is demonstrated that such a structured plate exhibits large band gaps due to the presence of heavy resonators linked to the host plate by tiny ligaments. In addition, the flat band representing the resonance frequency can be used to achieve new functionalities such as focusing. To do so, the ligaments should be tilted in order to facilitate the energy transfer of the twisting mode to the continuum, and vice versa. Since the platonic crystal is of constant thickness, the overall device could be implemented easily and might offer an alternative to free standing plates. Micro-Electro-Mechanical System applications could be targeted.

Y.A. thanks the ANR PLATON (No. 12-BS09-003-01) for financial support. S.G. and A. D. are thankful for the ERC funding through ANAMORPHISM Starting Grant.

\end{document}